\documentclass[traditabstract]{aa} 
\usepackage{graphicx,natbib,psfig,amssymb,multirow,sidecap} 
\newcommand{\ltsima} {$\; \buildrel < \over \sim \;$} 
\newcommand{\gtsima} {$\; \buildrel > \over \sim \;$} 
\newcommand{\lta} {\lower.5ex\hbox{\ltsima}} 
\newcommand{\gta} {\lower.5ex\hbox{\gtsima}} 
\newcommand{\Ha} {H$\alpha$}
\newcommand{\Hb} {H$\beta$}
\newcommand{\ergscm}{\>{\rm erg}\,{\rm s}^{-1}\,{\rm cm}^{-2}}
\newcommand{\ergs}{$\>{\rm erg}\,{\rm s}^{-1}$}
\newcommand{\kms}{$\rm{\,km \,s}^{-1}$}
\newcommand{\forb}[2]{\mbox{$[{\rm #1\, #2}]$}}
\newcommand{\oiii}{\forb{O}{III}}
\begin{document} 
\defcitealias{capetti11alias}{Paper~I}
\title{Revisiting the census of low-luminosity AGN.}
  
\titlerunning{Revisiting the census of low-luminosity AGN.}

\authorrunning{Alessandro Capetti}
  
\author{Alessandro Capetti \inst{1}} 
\offprints{A. Capetti}  
\institute{INAF - Osservatorio Astronomico di Torino, Via
  Osservatorio 20, I-10025 Pino Torinese, Italy \\
\email{capetti@oato.inaf.it}}

\date{}  
   
\abstract{The aim of this paper is to revisit critically the current census of
  AGN as derived from optical spectroscopy.

  We considered the spectra of nearby ($z<0.1$) galaxies from the Sloan
  Digital Sky Survey (SDSS). The equivalent width (EW) distribution of the
  [O~III]$\lambda5007$ emission line is strongly clustered around $\sim$0.6
  \AA, extending the validity of the results we obtained for red giant
  ellipticals. The close connection between emission lines and stellar
  continuum points to stellar processes as the most likely source of the bulk
  of the ionizing photons in these galaxies although their emission line
  ratios are similar to those of active nuclei. Genuine AGN might be sought
  mainly among the minority ($\sim$ 5-10 \%) of outliers, i.e., galaxies with
  EW$\gtrsim 2$\AA. The galaxies located in the AGN region of the
  spectroscopic diagnostic diagrams outnumber outliers by a factor 5-10 which
  casts doubts on the accuracy of the current identification of active
  galaxies, particularly those of LINERs of low line luminosity, $\lesssim
  10^{39}-10^{40}$ \ergs.

  This conclusion can be tested by using spectra that covers smaller physical
  regions such as those that are already available in the literature of the
  $\sim$ 500 nearest bright galaxies, with a stellar continuum reduced by a
  factor of 20-100 with respect to SDSS galaxies. If the emission lines were
  mainly of AGN origin, their contrast against the continuum should be
  enhanced. Instead, their EW distribution is similar to that of the SDSS
  sample, with just an increase of the outlier fraction.

  We conclude that the number low-luminosity AGN is currently largely
  overestimated with a sample purity as low as $\sim 10$\%. As a consequence
  the properties of low-luminosity AGN should be fundamentally revised.

  }

\maketitle

\section{Introduction}

Emission lines are among the most widely tools used to reveal an active
galactic nucleus (AGN) and to explore the properties of its central
engine. The luminosity of narrow emission lines not affected by nuclear
obscuration, is a robust estimator of the AGN bolometric luminosity
\citep{mulchaey96}. Diagnostic diagrams that compare emission line ratios can
distinguish H~II regions from gas ionized by nuclear activity
\citep{heckman80} and separate AGN into various sub-classes, e.g., Seyferts and
LINERs\footnote{Low ionization nuclear emission regions, \citep{heckman80}.}
\citep{kewley06}, possibly associated with different accretion modes
\citep{buttiglione10}. The analysis of extensive spectroscopic datasets can
then be used to obtain a detailed census of active galaxies, to study their
properties and to look for statistical connections with, e.g., the host
galaxies and environment.

In an initial analysis (\citealt{capetti11alias}, hereafter Paper I) we
studied the spectra of a large sample of nearby, red, giant early-type
galaxies (RGEs) from the Sloan Digital Sky Survey (SDSS).  We restricted the
analysis to galaxies with a Ca break strength $D_n(4000)>1.7$, with a stellar
velocity dispersion $\sigma_* > 156$ \kms (corresponding to a mass $M_*
\gtrsim 5 \times 10^{10} M_{\sun}$, \citealt{hyde09}) and retained only
early-type galaxies, i.e., objects with a concentration index $C_{r}\geq 2.86$
\citep{nakamura03,shen03}.

Most of these galaxies show emission lines in their spectra, e.g., the
[O~III]$\lambda5007$ emission line is detected in 53 \% of them. Furthermore,
the vast majority of RGEs for which it is possible to derive emission line
ratios (amounting to about half of the sample) show values characteristic of
LINERs. These results apparently lead to the conclusion that most of these
nearby luminous galaxies harbor an active nucleus.

\begin{figure*}
\centerline{
\includegraphics[scale=0.6,angle=0]{17937f1a.epsi}
\medskip
\includegraphics[scale=0.7,angle=0]{17937f1b.epsi}
}
\includegraphics[scale=0.95,angle=0]{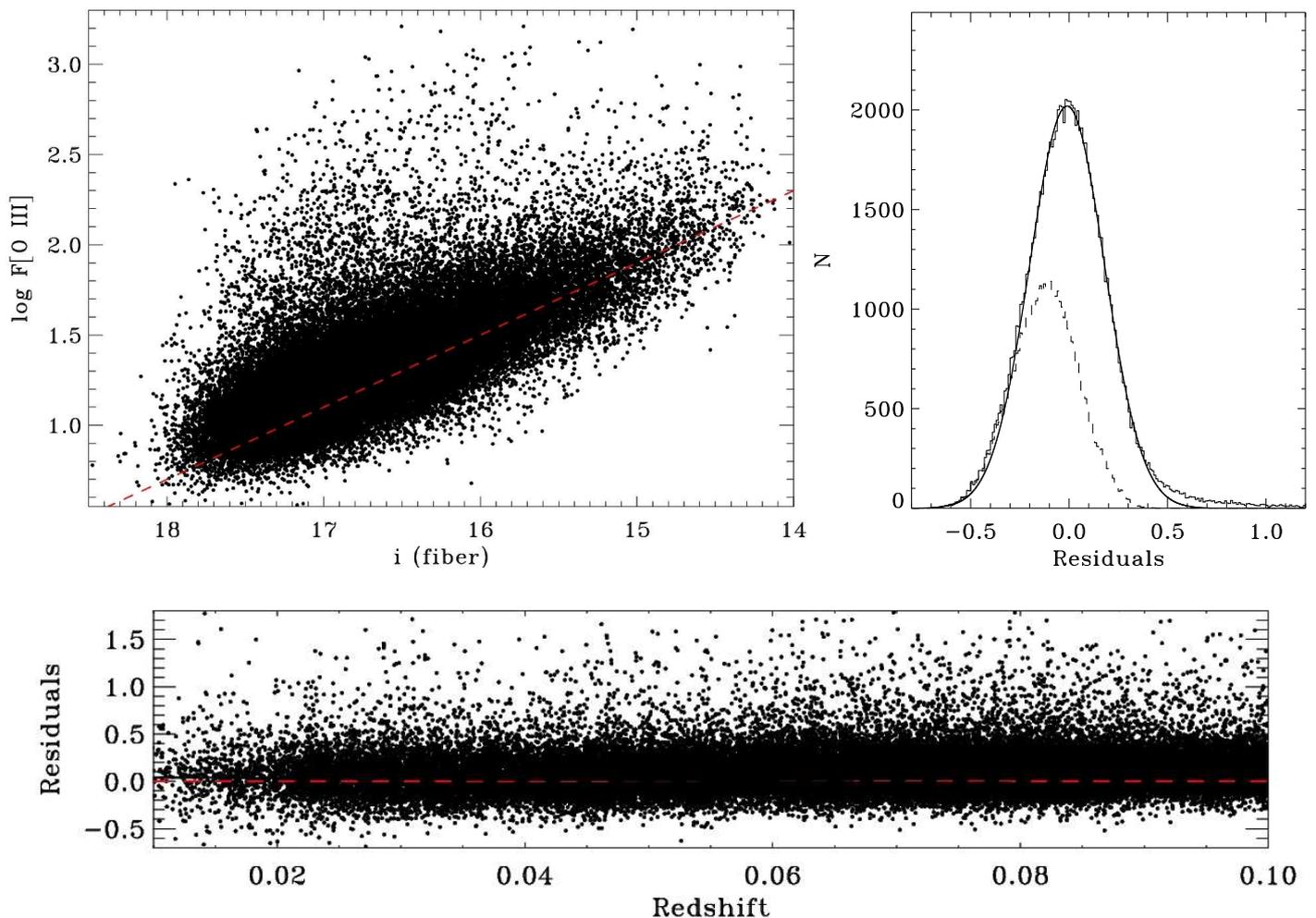}
\caption{Left panel: logarithm of the [O~III] emission line flux (in units of
  $10^{-17}\ergscm$) versus the k-corrected $i$-band magnitude within the SDSS
  fiber, both quantities corrected for galactic absorption. For clarity, upper
  limits in the line flux are not reported. The dashed red line corresponds to a
  constant ratio between the two quantities. Right panel: histogram of the
  residuals from the median line. The dashed histogram is the contribution of
  upper limits. The solid line
is a Gaussian distribution with a width of 0.19 dex. Bottom:
  residuals from the median vs redshift.}
\label{riga}
\end{figure*}

However, the [O~III] flux shows a strong correlation with the flux measured
within the 3$\arcsec$ SDSS fiber in the $i$-band. The $i$-band magnitude
(being less affected by uncertainties in the k-correction, absorption, and by
differences in the stellar population) is a good estimator of the stellar mass
within the fiber. Thus the ratio between lines and stellar mass is essentially
constant (showing a dispersion of only 0.18 dex).  Furthermore, there is no
significant change of EW with redshift, despite the change in the size of the
region covered by the SDSS fiber from $\sim$0.6 to 5.5 kpc.  The close
connection between emission lines and stellar continuum points to stellar
processes as the most likely source of the bulk of the ionizing photons in
RGEs, rather than active nuclei. In particular, it has been suggested
  that hot evolved stars can play a dominant role
  (e.g., \citealt{trinchieri91,binette94}), an idea also supported by the
  results obtained from photoionization models that are able to reproduce the
  observed EW and optical line ratios (e.g., \citealt{stasinska08,sarzi10}).
Shocks driven by supernovae or stellar ejecta might also contribute to the
ionization budget. Conversely, these results are not straightforward to
explain if the emission lines are powered by an AGN. 

A minority, $\sim$4\%, of the galaxies show emission lines with an equivalent
width a factor of $\gtrsim 2$ greater than the sample median. Seyfert-like
spectra are found only among these outliers. Furthermore, 40\% of this
subgroup have a radio counter part, against $\sim$6\% of the rest of the
sample. These characteristics argue in favor of an AGN origin for their
emission lines. However, emission line diagnostic diagrams do not reveal a
distinction between the AGN subset and the other members of the sample, and
consequently they are not sufficient to establish the dominant source of the
ionizing photons, which is better predicted by the EW of the emission lines.

In this paper we relax the selection criteria, expanding our analysis to all
nearby galaxies with spectra available from the SDSS Data Release (DR) 7.
While in Paper I we focused only on giant elliptical galaxies, in Section
\ref{old} we consider all spectra with a Ca break strength $D_n(4000)>1.7$,
i.e., all galaxies for which the SDSS fiber covers an old stellar population,
dropping the requirements on mass and morphology.
In Section \ref{all} we move to the analysis of all SDSS DR7 spectra of nearby
galaxies, regardless of their spectro-photometric properties. In order to
improve the AGN census to the lowest luminosities, in Section \ref{ho} we also
consider the 486 nearby galaxies studied spectroscopically by \citet{ho95}
with the Hale telescope at Mount Palomar.

We discuss our results in Section \ref{discussion} and present our summary and
conclusions in Section \ref{summary}.

\section{Emission lines properties of red galaxies}
\label{old}

We here consider the $\sim$ 100,000 SDSS \citep{york00} Data Release (DR) 7
spectra characterized by a Ca break strength $D_n(4000)>1.7$ (and an error $<
0.05$)\footnote{This corresponds effectively to the inclusion only of spectra
  with a median signal-to-noise ratio higher than $\sim 10$.}, associated with
$\sim$ 90,000 unique galaxies at $z<0.1$ based on the MPA-JHU DR7 spectrum
measurements, available at {\sl
  http://www.mpa-garching.mpg.de/SDSS/DR7/}. Then, we are studying all nearby
galaxies for which the SDSS fiber covers an old stellar population. For the
sake of brevity we define them as `red galaxies'. Most of them are early-type
galaxies, now spanning their full range of mass, but there is also a
substantial number of spiral galaxies where the bulge is sufficiently extended
to fill the SDSS fiber.

\subsection{Emission lines and stellar continuum}
\label{el}

\begin{figure}
\centerline{
\includegraphics[scale=0.5,angle=0]{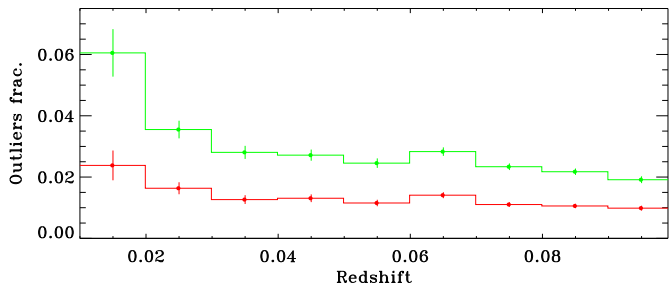}
}
\caption{Fraction of outliers versus redshift for the red galaxies, dominated
  by an old stellar population, i.e., with $D_n(4000)>1.7$. Green dots and
  histogram are weak outliers, the red ones are strong outliers.}
\label{out-z}
\end{figure}

We considered the [O~III]$\lambda5007$ emission line flux measured after
subtraction of a starlight template (see \citealt{kauffmann03b} for a detailed
description of the method used for the continuum subtraction). The [O~III]
line is detected at a significance higher than 3$\sigma$ in $\sim$ 53\% of the
red galaxies. Fig. \ref{riga} compares the [O~III] flux with the k-corrected
$i$-band magnitude within the SDSS fiber, both quantities corrected for
galactic absorption. The strong connection found in Paper I is confirmed with
the vast majority of the red galaxies clustered in a narrow stripe. Slope and
intercept of the best-fitting logarithmic linear relation are
indistinguishable from those derived for RGEs.

\begin{figure*}
\centerline{
\includegraphics[scale=0.9,angle=0]{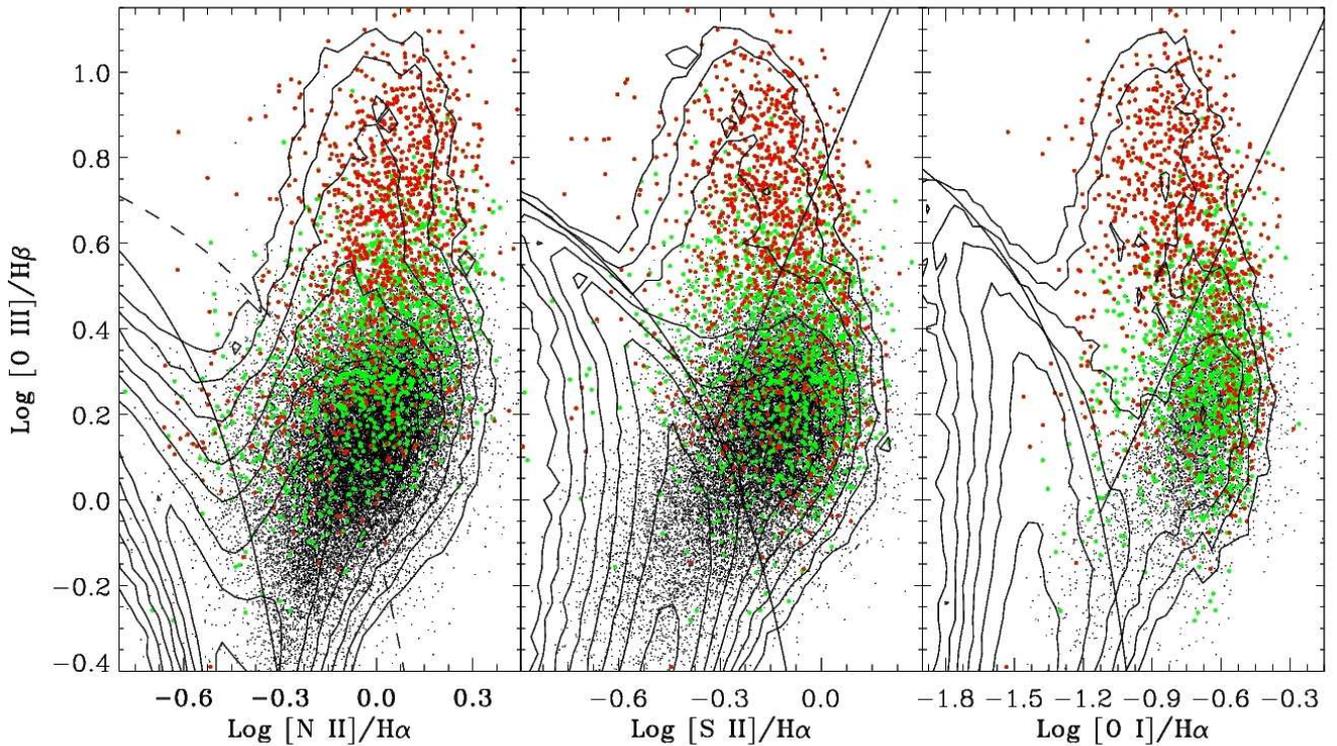}
}
\caption{Spectroscopic diagnostic diagrams for the red galaxies. The solid
  lines are from \citet{kewley06} and separate star forming galaxies, LINER,
  and Seyfert; in the first panel the region between the two curves is
  populated by the composite galaxies. Small black dots mark the location of
  all red galaxies, while the large colored dots mark the outliers,
  i.e.,  objects with a line excess with respect to the median value by a
  factor of R\oiii$>$5 (red) or 3 (green) in Fig. \ref{riga}. Contours
  represent the iso-densities of all DR7 emission line galaxies with $z<0.1$.}
\label{diag}
\end{figure*}

In the right panel we show the histogram of the residuals from the median line
(the dashed histogram is the contribution of upper limits) that is well
reproduced by a Gaussian distribution with a dispersion of only 0.19 dex
(similar results, although with slightly broader distributions, are obtained
using different lines and/or continuum bands). Nonetheless, there are outliers
from this trend. Setting a threshold at R\oiii\footnote{R\oiii\ is defined as
  the ratio between the \oiii\ flux of a galaxy with respect to its median
  value at a given magnitude.} = 5 we find 1.2\% of `strong' outliers, a
fraction that increases to 2.5\% of `weak' outliers when lowering the limit to
3. \footnote{As a comparison, the fraction of outliers for RGEs is 0.5\% for
  R\oiii $>$ 5 and 1.2\% at R\oiii $>$ 3.}

A more readily measurable parameter is the EW of the [O~III] line. For the
SDSS galaxies it has been estimated by considering the continuum level 200
pixels around the line. The EW distribution is also highly concentrated
with a median value of $\sim 0.6$ \AA\ (including also
upper limits in the analysis) and a dispersion of 0.21 dex.

We looked for trends between the residuals and the spectrophotometric
parameters used for the sample selection (namely redshift and $D_n(4000)$) but
we failed to find any statistically significant link. There is instead a weak
trend for an increase of the fraction of outliers moving from higher to
lower redshifts (see Fig. \ref{out-z}) 
particularly clear in the two bins at the lower redshifts.

As we already discussed in \citetalias{capetti11alias}, because we are studying
emission lines that are generally of very low EW, it is important to assess the
reliability of these measurements and of the error estimates. The comparison of
the results obtained from duplicated observations of the same galaxies
indicates that the assessment of the statistical errors provided by the SDSS
database is robust. Indeed the differences between pairs of measurements
closely match those predicted by their uncertainties. The good agreement of
the flux measurements obtained by \citet{oh11} using a different approach for
the removal of the stellar emission also supports the robustness of those
provided by MPA-JHU database.  

We also considered the possibility of an additional error related to the
accuracy of the subtraction of the stellar emission (see
\citealt{annibali10}) by adding in quadrature a constant error in EW of 0.16
\AA\ for \Hb\ and of 0.08 \AA\ for the other key emission lines. Our results
on the connection between \oiii\ flux and $z$ magnitude are only marginally
changed by adopting this more conservative error treatment. The fraction of
detected sources decreases from 53 to 49\%, but the slope of the relation
remains unchanged. The number of outliers is not affected. We anticipate that,
similarly, the results on the spectroscopic diagnostic diagrams, presented in
the next section, are not substantially affected; the main difference is a
reduced fraction of galaxies that can be classified, but their general
location is not changed.

\begin{figure}
\centerline{
\includegraphics[scale=0.48,angle=0]{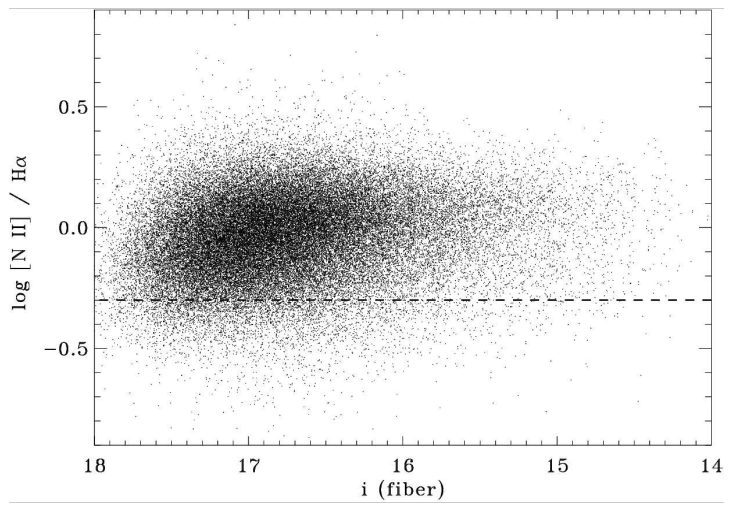}
}
\caption{[N~II]/\Ha\ ratio for all red galaxies with both lines
  detected. Emission lines with a ratio that locate an object above the
  horizontal dashed line are unlikely to be powered by star formation.}
\label{n2ha}
\end{figure}

\subsection{Spectroscopic diagnostic diagrams}

Fig.~\ref{diag} shows the location in the spectroscopic diagnostic diagrams
(e.g., \citealt{heckman80,baldwin81,veilleux87,kewley06}) for the selected
galaxies that have all relevant emission lines detected at SNR$>$3 separately
for each diagram. Starting from the left side, the percentages with respect to
the whole sample are 30, 25, and 8\% in the three diagrams,
respectively. The majority of the objects fall in the LINERs region, while the
Seyfert and star-forming regions are scarcely populated. In particular
(according to their location in the left diagram) only 0.6\% of the sample
have line ratios typical of star-forming galaxies, 10.8\% are the composite
galaxies, while 18.9\% of the red galaxies are in the AGN region.

With respect to the overall population, our selection filters out most of the
objects at (or close to) the location of star-forming galaxies. This can be
explained as the consequence of the exclusion of the objects with a
substantial population of young stars whose emission lines have a significant
contamination from star-forming regions.

Approximately $\sim$70 \% of the galaxies have no optical classification
because at least one of the key emission lines is undetected. Nonetheless, in
85\% of them, the [N~II] and \Ha\ lines are both detected; this allows us to
derive at least a crude classification. Indeed, log([N~II]/\Ha) $> -0.3$ in
70 \% of them (see Fig. \ref{n2ha}), a threshold above which no star-forming
galaxies are found. On the other hand, Seyfert galaxies have usually bright
emission lines, therefore they are expected to be all properly cataloged by the
diagnostic diagrams in Fig.~\ref{diag}. This leads to the conclusion that
these galaxies, which amount to more than half of the sample, generally
are LINERs.

Let us now consider the properties of outliers, i.e., the objects with EW
exceeding by a factor of at least 3 the median of the sample. `Strong'
outliers (i.e., the objects with R\oiii\ $>$ 5) essentially represent the
totality of Seyferts, but many of them are LINERs, while weaker outliers (with
$3 <$ R\oiii $< 5$) are mostly LINERs. It is important to note that they
cannot be readily separated from the bulk of the red galaxies population.

We conclude that the main results obtained in Paper I are confirmed by the
analysis of all `red galaxies', regardless of their mass, suggesting that the
bulk of red emission line galaxies are not genuine AGN. While $\sim$ 20\% of
the red galaxies are located in the AGN region of the diagnostic diagrams (and
the fraction of LINERs is probably as high as $\sim$ 70\% based on the
[N~II]/\Ha\ ratio), the majority of them have emission lines most likely
powered by stellar processes. Genuine AGN might be sought mainly among the
$\sim$ 1-2 \% of outliers, i.e., galaxies with higher EW values. The number of
AGN hosted by red galaxies derived from the emission line ratios is probably
overestimated by a factor ranging from 10 up to 70.

\begin{figure}
\centerline{
\includegraphics[scale=0.85,angle=0]{17937f5.epsi}}
\caption{Distribution of the residuals from the median line for all nearby DR7
  galaxies. The dashed histogram is the contribution of upper limits. The
  solid line represents a Gaussian distribution with a width of 0.20 dex.}
\label{res-hist-all}
\end{figure}

\begin{figure*}
\centerline{
\includegraphics[scale=0.9,angle=0]{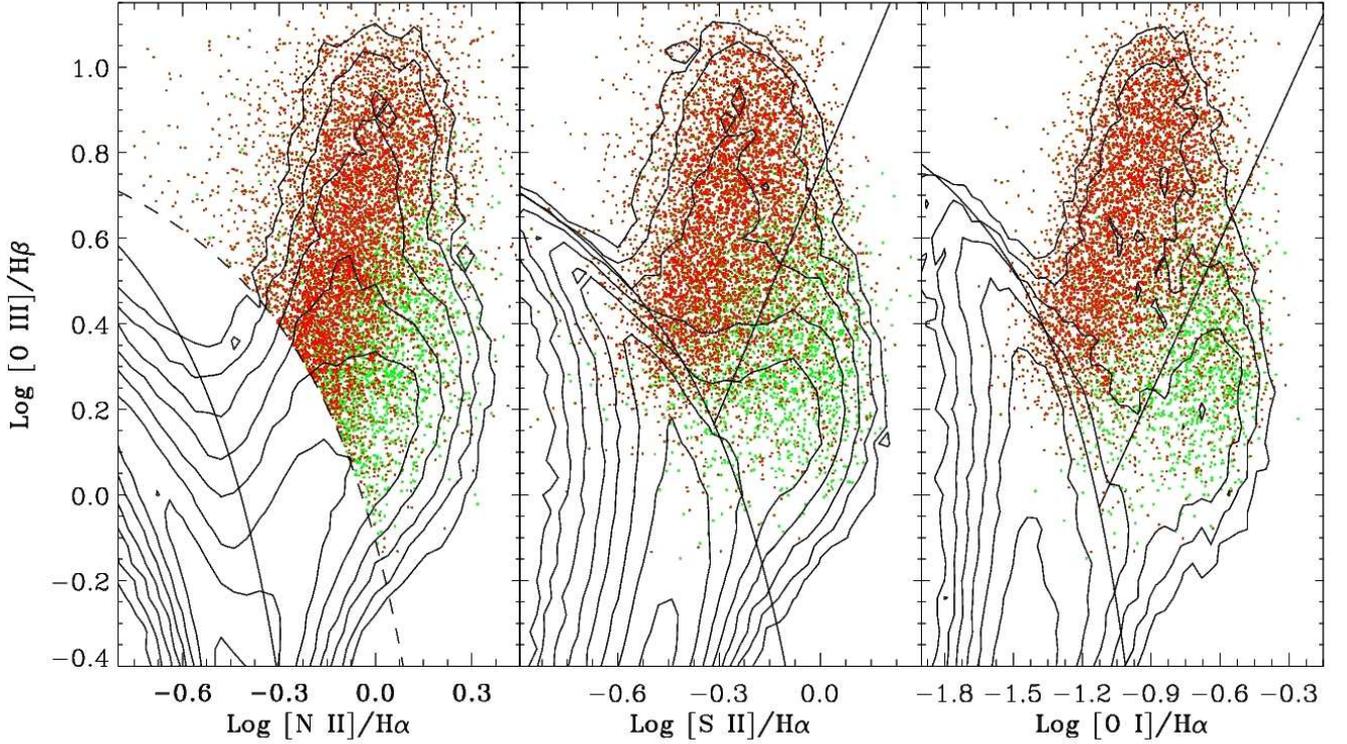}
}
\caption{Spectroscopic diagnostic diagrams for all DR7 emission line galaxies
  with $z<0.1$. Contours represent their iso-densities and are drawn with a
  common ratio of 2. Dots show the location of the outliers (i.e., objects with
  a line excess with respect to the median value by a factor of R\oiii$>$5
  (red) or 3 (green) in Fig. \ref{riga}) limiting to those falling in the AGN
  in the left diagram, i.e., excluding star-forming and composite galaxies.}
\label{kewley1}
\end{figure*}

\begin{figure}
\centerline{
\includegraphics[scale=0.5,angle=0]{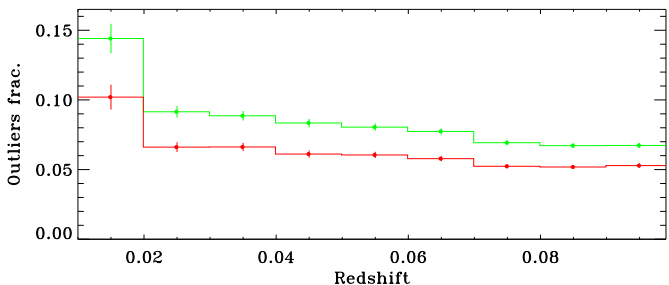}
}
\caption{Fraction of outliers versus redshift for all galaxies with
  $z<0.1$ and located in the AGN region in the left panel of Fig. \ref{kewley1}. The weak histogram is for weak outliers, the red one for strong
  outliers.}
\label{out-z-all}
\end{figure}

\begin{figure*}
\centerline{
\includegraphics[scale=0.5,angle=0]{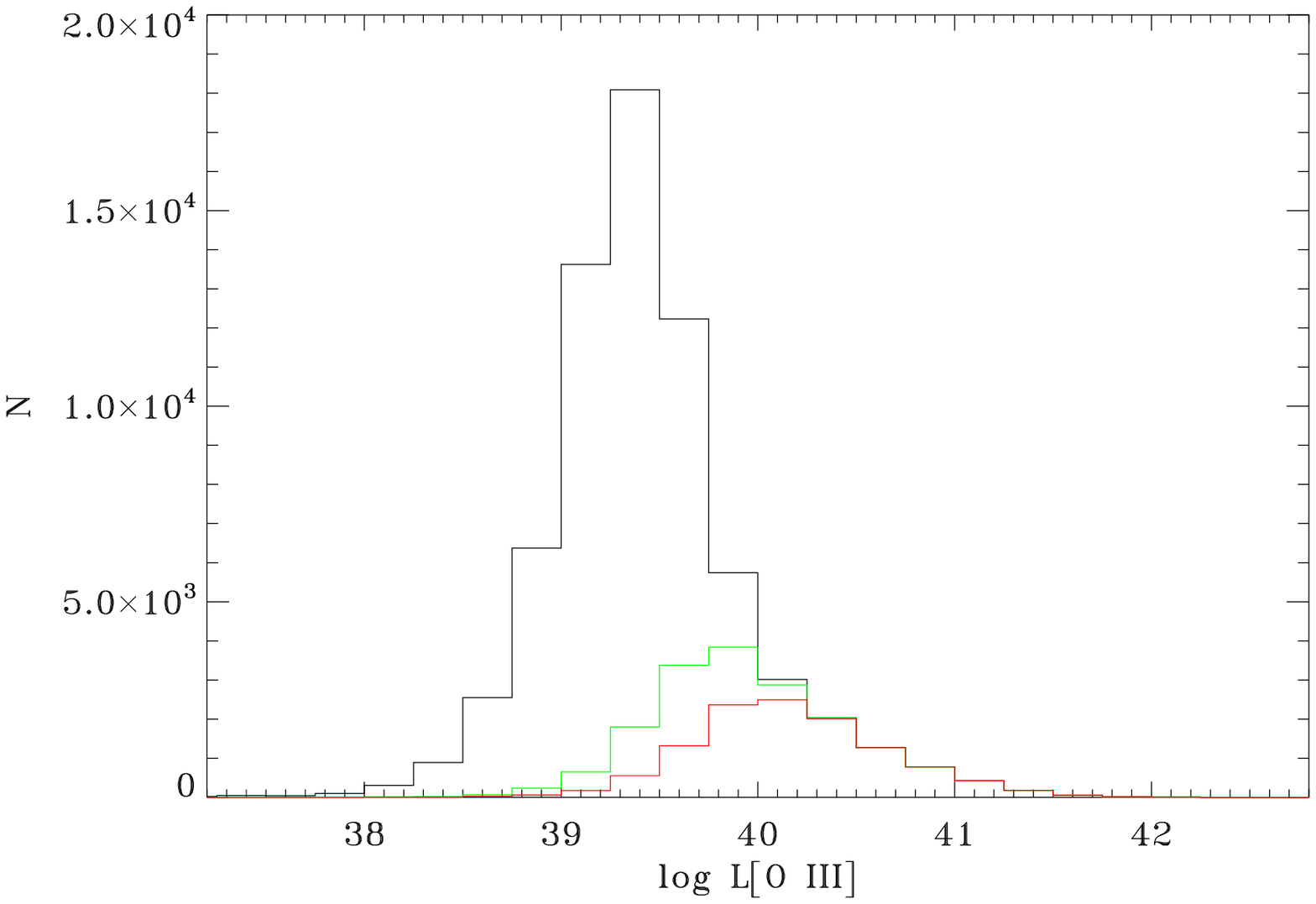}
\includegraphics[scale=0.5,angle=0]{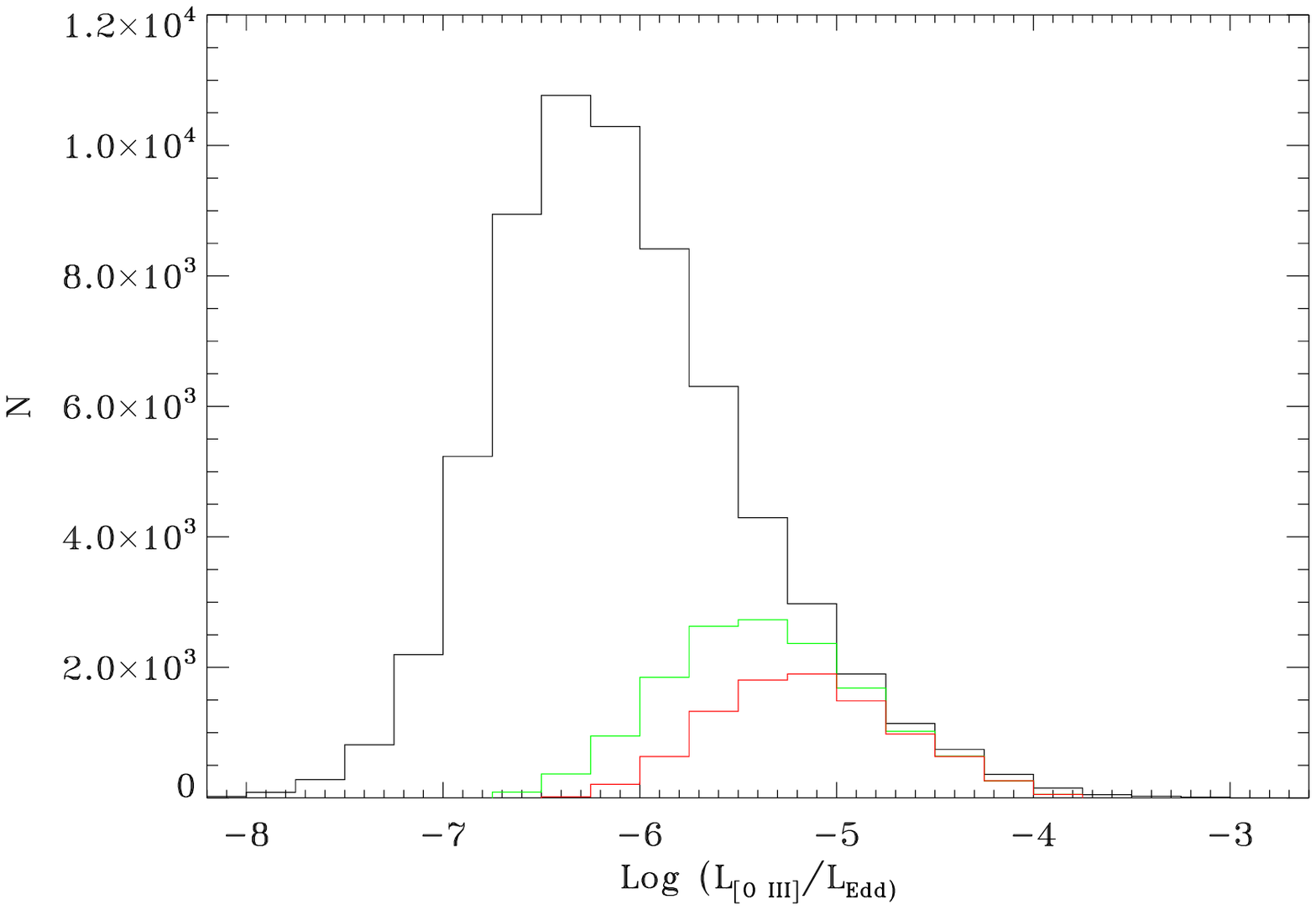}
}
\caption{Left: histogram of the [O~III] luminosity (in \ergs) for all galaxies
  located in the AGN region in Fig. \ref{kewley1}; the contribution of strong
  (weak) outliers is shown in red (green). Right: same as the left panel, but
  with the [O~III] luminosity measured in fraction of the Eddington
  luminosity.}
\label{lo3}
\end{figure*}

\section{Emission line properties of nearby galaxies}
\label{all}
We now also drop the requirement of a large Ca break,
but, for consistency, we maintain the requirement that its error must be
smaller than $< 0.05$ and the restriction on redshift ($z<0.1$).  This
selection yields $\sim$315,000 spectra.

Because we are mainly interested in the study of AGN we excluded from the
analysis the $\sim$155,000 galaxies spectra with emission line ratios that
locate them in the diagnostic diagram in the region characteristic of star
forming and, at least in a first stage, also the $\sim$39,000 composite
galaxies.

The \oiii\ line is detected in 46\% of the remaining $\sim$121,000 spectra and
we recovered the strong connection between line flux and stellar continuum.  The
distribution of the residuals from the median line (the same derived in
Sect. \ref{el}) is still highly concentrated, see Fig. \ref{res-hist-all}. The
peak of the distribution is well described by a Gaussian with a width of 0.20
dex and shifted with respect to that reproduced in Fig. \ref{riga} by only
0.01 dex.  However, there is a more prominent tail of galaxies with higher
ratios, with 5.8\% of outliers with R\oiii$>$5 (7.6\% for R\oiii$>$3), i.e., 3
- 5 times more than for the red galaxies.

There is again a weak increase of the fraction of outliers from higher to
lower redshifts (see Fig. \ref{out-z-all}), but less pronounced than
that observed for the red galaxies. Here there are $\sim$ 7\% outliers at
$z~\sim0.1$, increasing to $\sim$8\% for $z~\sim0.02-0.03$, and only in the
first bin ($z~\sim0.01-0.02$) the fraction reaches $\sim$14\%.

In the diagnostic diagrams we can locate $\sim$ 24\% of these
galaxies. Because we excluded all star-forming and composite galaxies, these
all lie in the AGN region in Fig. \ref{kewley1}. Strong outliers are mostly
located in the Seyfert region, while the weak ones are generally LINERs.
Furthermore, in 65 \% of them we measure log([N~II]/\Ha) $> -0.3$, which
argues in favor of a LINER classification.

The selection of outliers based on their higher EW also induces a separation
between them and the rest of the galaxies of the sample in terms of line
luminosity, L$_{\rm[O~III]}$. Fig. \ref{lo3} shows the distributions of
L$_{\rm[O~III]}$ of the sources located in the AGN region in
Fig. \ref{kewley1} separately for the galaxies of low EW and for the
outliers. Strong outliers have a median luminosity of $10^{40}$ \ergs, $\sim$6
times higher than the rest of the sample. If we refer L$_{\rm[O~III]}$ to the
Eddington luminosity (adopting the \citealt{tremaine02} relation between black
hole mass and stellar velocity dispersion) the ratio between the two medians
increases to a factor of $\sim$20.

If we adopt a less conservative AGN definition, i.e., not excluding the
composite galaxies, the objects with a \oiii\ detection are 59\%. The results
in terms of the distribution of the residuals are similar, but with a slightly
higher fraction of outliers (7.6\% and 11.5\% at R\oiii$>$3 and 5,
respectively); those that can be located in a least one of the diagnostic
diagrams are instead 42\%, and 67\% of them have log([N~II]/\Ha) $> -0.3$.

Therefore, again, although the fraction of outliers is significantly higher
than for the `red galaxies', most of the nearby galaxies are likely to have
emission lines powered by stellar processes. The ratio between
objects with line ratios typical of AGN and outliers (the likely genuine AGN)
is still very high, a factor of $\sim$ 5 - 10.

\section{Revisiting the Palomar survey}
\label{ho}

Because the dominant contamination to the emission lines is apparently
associated with stellar processes, significant progress for a proper AGN
census can be obtained considering spectra obtained in a smaller physical
region, with a reduced contribution from the stellar emission. These data are
already available in the literature from the spectroscopic survey performed at
the Palomar observatory of 486 nearby (with an average recession velocity
of $\sim 1000$ \kms) bright galaxies \citep{filippenko85}. The objects
of the `Palomar sample' have a median distance $\sim$ 20 times smaller than
the SDSS galaxies and as consequence the emission associated with AGN of lower
luminosity may be visible in their spectra.

\begin{figure*}
\centerline{
\includegraphics[scale=1.,angle=0]{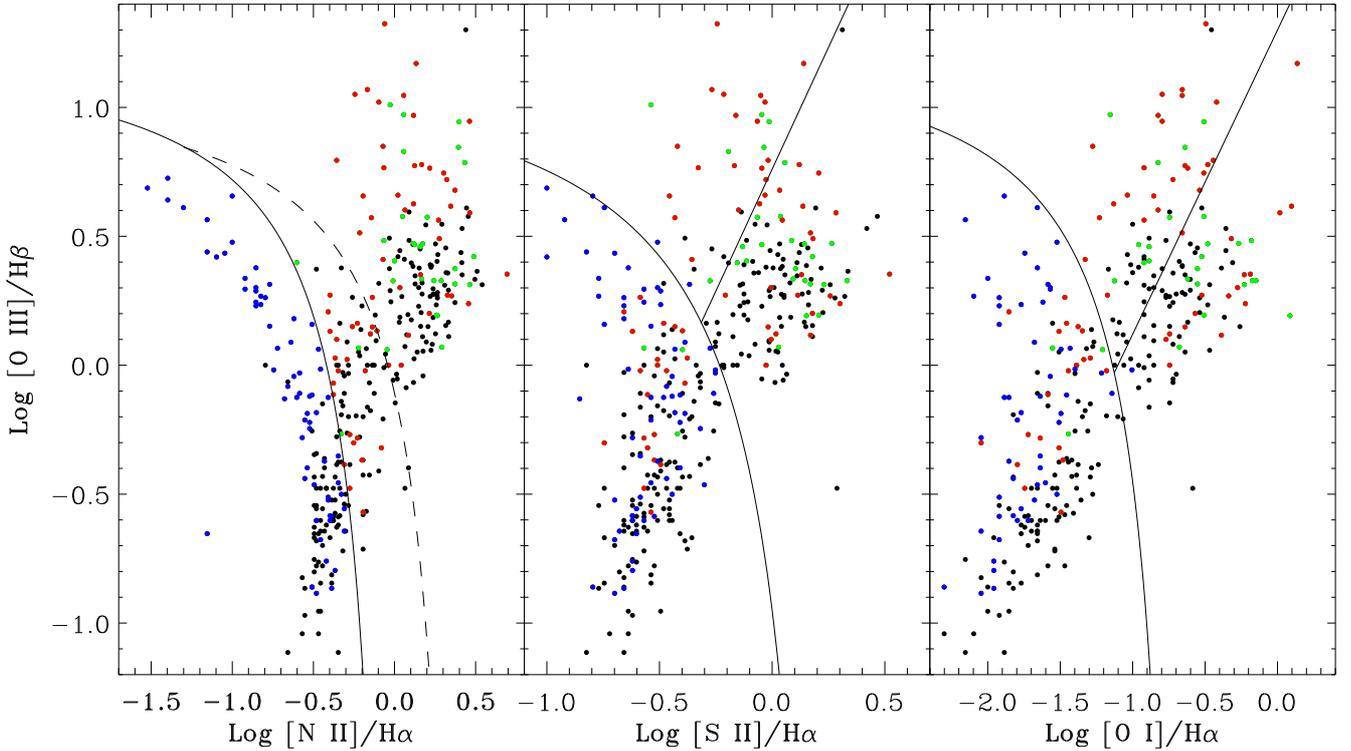}
}
\caption{Spectroscopic diagnostic diagrams for the Palomar sample, adapted
  from \citet{ho97}. Black dots mark the location of all galaxies of low EW
  (see text for details). Colored dots mark the outliers, i.e.,  objects with a
  line excess with respect to the median value measured for `red galaxies' by
  a factor of R\oiii$>$5 (red) or 3 (green). Blue dots are
  the star-forming galaxies according to their location in the left panel.}
\label{diag-ho}
\end{figure*}

Optical spectra were obtained at the Hale 5 m telescope at Palomar Observatory
for almost all bright galaxies in the northern sky, selected from the Revised
Shapley-Ames Catalog of Bright Galaxies \citep{sandage81} with a limit of
B$_{\rm T}<$12.5 \citep{ho95,ho97}. 
A 2$^{\prime\prime}$ slit was generally used and the spectra were extracted
over a synthetic aperture of 4$^{\prime\prime}$x2$^{\prime\prime}$.
Coincidentally, this extraction region has the same size as the circular,
3$\arcsec$ in radius, fiber of the SDSS.

Only 68 galaxies turn out be lineless, while 418 objects are emission line
galaxies.\footnote{From this initial list we excluded a few sources not
  formally part of the sample and included for historical reasons, mainly AGN
  with a negative declination.} It is possible to locate a very high part
of these galaxies in the diagnostic diagrams (see Fig. \ref{diag-ho}), between
66 and 76\%, depending on the pair of ratios considered. Based on their
position in the left diagram, 38 \% of them are AGN, 15 \% are composite
galaxies, and 24 \% are star-forming galaxies.\footnote{The remaining galaxies
  are 14 \% lineless and 9 \% that have at least one of the 4 relevant
  emission lines undetected.}

\citet{ho97} have listed the EW values for the \Ha\ line; we used these data
to measure the ratio between the \oiii\ and the continuum around the \Ha\
line. Although this is not a conventional measurement, we can estimate this
same parameter for the red galaxies discussed in Sect. \ref{el} for reference.

\begin{figure}
\centerline{
\includegraphics[scale=0.5,angle=0]{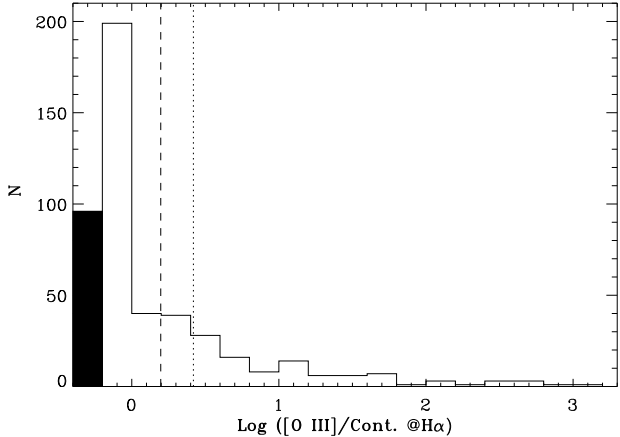}}
\caption{Logarithm of the ratio between the \oiii\ flux and the continuum
  around the \Ha\ line for the Palomar sample. The vertical lines are located
  at 3 and 5 times the median value measured for the SDSS red galaxies,
  i.e., 0.52 \AA. The first bin of the distribution contains the 28 emission
  line galaxies where the \oiii\ line is not detected and the 68 lineless
  galaxies.  }
\label{ewhist-ho}
\end{figure}

The distribution of EW$_{\rm {[O~III] @ H}\alpha} = {\rm
  log}\,F_{\rm[O~III]}/F_{\rm{cont,H_\alpha}}$ of the SDSS red galaxies is
well reproduced by a Gaussian centered at 0.52 \AA.  The distribution of
EW$_{\rm {[O~III] @ H}\alpha}$ for the Palomar sample is shown in
Fig. \ref{ewhist-ho}. The first two bins (including galaxies with values $<$ 1
\AA, the 28 emission line galaxies where the \oiii\ line is not detected and
the 68 lineless galaxies) contain more than 60\% of the sample. Although the
statistics is more limited than for the SDSS galaxies, the emission line
galaxies in the Palomar sample are also mainly objects of low equivalent
width.

In particular, only 1/3 of the 175 galaxies located in the AGN region of the
diagnostic diagrams have EW$_{\rm {[O~III] @ H}\alpha}> 3
\times 0.52$ \AA, reducing the fraction of likely active nuclei from 38\% to
12\% (and there are only 5\% of composite galaxies). Moving the threshold to 5
$\sigma$, there are 7\% AGN and 4\% composites.

The results obtained from the diagnostic diagrams built considering
[S~II]/\Ha\ and [O~I]/\Ha\ ratios are similar: the percentage of AGN decreases
from $\sim$32-35 \% to $\sim$12 \% (and 8\% at 5 $\sigma$). This is mostly
owing to a strongly reduced number of LINERs, because less than 20 \% of them
have an equivalent width higher than the adopted threshold. In contrast, only
a minority of (possible) Seyferts have a low EW and they are generally located
close to the boundary with LINERs.

The line luminosity of the Palomar galaxies located in the AGN region is a
factor of $\sim$ 10 lower than the SDSS galaxies, reaching values as low as
L$_{\rm[O~III]} \sim 10^{37}$ \ergs. The outliers form the bright end of this
distribution and have a median of L$_{\rm[O~III]} \sim 10^{39}$ \ergs.

\section{Discussion}
\label{discussion}
The spectra of nearby ($z<0.1$) galaxies extracted from the SDSS DR7 show that
even after removing the star-forming galaxies, most of the galaxies are
emission line galaxies. The percentage of objects falling into the AGN region
of the diagnostic diagrams is $\sim$24\%, generally showing line ratios
typical of the subcategory of LINERs. Considering the objects for which an
identification can be obtained from the [N~II]/\Ha\ ratio as well, this
percentage is increased to $\sim$ 65\%.  Apparently, this is an indication
that most of these galaxies host an active nucleus.

However, their distribution of line equivalent widths is strongly clustered
around EW$_{\rm [O~III]}\sim 0.6$\AA. This result is very difficult to account
for if the emission lines are powered by an AGN because this requires a
fine-tuning between the strength of the nuclear ionizing field, the spatial
distribution of the emission lines, and the stellar mass. Conversely, the
strong connection between line emission and stellar continuum points to a
stellar origin of the emission lines. As already noted in the introduction,
the observed EW and optical line ratios are consistent with the predictions of
photoionization from hot evolved stars.  Only the minority of galaxies
($\lesssim 10\%$ of the sample) that does not obey this trend, because they
have higher EW, should be considered as genuine AGN.

An analogy can be drawn between the origin of emission lines and the source of
dust heating: recent Herschel observations of M~81 and M~33
\citep{bendo10,boquien11} show that the diffuse photon field produced by the
evolved stellar population plays a fundamental role in setting the dust
temperature, competing against young stars and (in the case of M~81) the
active nucleus in setting the energy budget. 

A test to probe the dominant role of old stars can be performed considering
objects located at different distances by only assuming that the galaxy's
population does not vary with redshift (a point that we will consider in more
detail below). If the emission lines were of nuclear origin, their contrast
against the stellar continuum should be enhanced in closer objects, leading to
an increase of the lines EW, because the fixed angular size of the fiber
covers a smaller portion of the galaxy. Conversely, in case of a dominant
stellar origin, lines are produced over an extended region and no significant
changes are expected in the distribution of equivalent width with
redshift. This is the result of our analysis, because the peak of the
distribution of the ratio between line flux and continuum does not vary
significantly with redshift.

The Palomar survey provides us with the data required to perform an even more
stringent test because it includes objects at an average distance $\sim$ 20
times smaller than the SDSS galaxies (their average redshift is
$z=0.063$). Although the average surface brightness within the extraction
region for the Palomar sources is $\sim$2-3 magnitudes higher than for the
SDSS sample, this only partly compensates for the strongly reduced physical
size of the extraction region, which leads to a typical decrease of the
stellar continuum by a factor of 20-100. However, the distribution of the
ratio between line and continuum still shows a very strong concentration at
low values, with more than 60\% objects having values $<$ 1 \AA, confirming
that the change in the stellar continuum is closely coupled with a decrease in
the line flux. This again points toward the conclusion that most of the line
emission originate from an extended region.

Nonetheless, a change with redshift is observed in the fraction of outliers
that is slightly higher in closer SDSS galaxies, increasing from an average of
7.6\% to 14\% for the objects with $z~\sim0.01-0.02$. This result can be
reconciled with the stable location of the peak of the EW distribution if, in
addition to the dominant diffuse line emission, there is also a contribution
from a nuclear emission line region. In the more distant objects, this is
generally swamped by the stellar emission. This might be sufficient to
increase the lines EW within the aperture in nearer galaxies at the same
level of luminosity and give rise to an outlier. This is confirmed by a
further increase of likely active galaxies among the Palomar galaxies, where
one third of the objects in the AGN region also show a high EW, twice the
fraction of the nearer SDSS galaxies.

The distribution of the \oiii\ luminosity of the outliers indicates that we
can currently isolate AGN in SDSS galaxies down to L$_{\rm[O~III]}\sim10^{39}$
\ergs, but most of them are 10 - 100 times brighter. This limit decreases
by a factor of $\sim$ 10 for a few Palomar galaxies, but even in this sample
the bulk of the AGN have L$_{\rm[O~III]} \gtrsim 10^{39}$ \ergs.

In principle, the information on the variation of the fraction of likely AGN
with redshift could be used to statistically model the luminosity function of
lower-luminosity AGN. However, as noticed above, the comparison between the
galaxies from the SDSS and Palomar surveys (but even with the same survey but
at the different redshift) should be treated with some care. Indeed, the
selection criteria of the two surveys are different and, within the SDSS
sample, the requirements to include a galaxy in the spectroscopic database
\citep{strauss02} causes a change of the target properties with redshift. If
there is a relation between the line emission and host properties,
particularly in case of an AGN origin, this might have an impact on the
fraction of outliers at various redshift. For example, several local low-mass
galaxies are included in the Palomar sample that would not enter in the SDSS
spectroscopic dataset were they located at the median SDSS redshift. While the
SDSS galaxies have an absolute magnitude mostly in the range M$_{\rm r}=-$20 -
$-$23, in the Palomar sample $\sim$ 15\% of the galaxies do not fall into this
range, because they have a lower luminosity. This low fraction does not alter
our general result significantly, even in the assumption that i.e., low-mass
galaxies do not host active nuclei, but a more detailed comparison requires a
careful matching of the samples.

\medskip

Our results point to the conclusion that the census of AGN based on the
optical spectroscopic diagnostic diagrams currently includes a dominant
percentage of objects (possibly as high as 90 \%) where the emission lines are
not powered by an active nucleus. This is particularly severe for the lowest
luminosity AGN, which remain virtually unexplored with the currently available
data. This clearly has strong implications for our understanding of the origin
and evolution of AGN. Indeed, a statistical study based on the analysis of
sample whose purity is only at the $\sim$ 10\% level necessarily leads to
flawed results.  The situation is slightly better for the nearer Palomar
galaxies, but in this sample as well only one third of the objects usually
classified as AGN are likely to be genuine active galaxies.

As a consequence the properties of low-luminosity AGN, e.g., their connection
with the host galaxies and environment, should be fundamentally revised. This
problem is particularly severe for objects of low line luminosity ($\lesssim
10^{39}-10^{40}$ \ergs) with line ratios typical of LINERs.

\section{Summary and conclusions}
\label{summary}

In Paper I we showed that most nearby ($z<0.1$), red giant early-type
galaxies are emission line galaxies and generally show emission line ratios
characteristic of LINERs. Apparently, this is an indication of a very large
percentage of AGN among these sources. 

However, their [O~III] flux shows a strong correlation with the flux measured
within the 3$\arcsec$ SDSS fiber in the $i$-band, a robust estimator of the
stellar mass covered by the fiber used to obtain the spectra.  This result is
very difficult to explain if the emission lines are powered by an AGN. 
Conversely, it suggests that at the origin of the observed emission
lines there are processes related to the stellar population. For example, the
observed EW and optical line ratios are consistent with the predictions of
models in which the photoionization is caused by hot evolved stars. As a
consequence, only a minority of these emission line galaxies is likely to be
powered by an active nucleus and this requires to revisit the current census
of AGN, particularly for those of lower luminosity.

We here extended our analysis to all nearby ($z<0.1$) SDSS galaxies.
Initially we considered all galaxies where the fiber covers an old stellar
population characterized by a calcium break strength greater than
$D_n(4000)>1.7$, but removing any other constraint based on morphology or
mass. As a consequence we included in the analysis early-type galaxies of any
mass, as well as spiral galaxies (provided that their bulges are sufficiently
extended to fill the fiber). The results obtained are effectively
indistinguishable from those we found limiting our search to RGEs. The strong
correlation between emission line fluxes and stellar continuum is confirmed
with the same slope and intercept (within 0.03 \AA). The distribution of the
residuals from the median value is again found to show a dispersion of only
0.19 dex. The number of outliers, i.e., of the objects with line flux higher
by a factor of at least 3 with respect to the median of the sample, is
slightly higher. We found 1.2\% galaxies with an line excess higher than 5
(against 0.8\% for RGEs), and 2.5 \% for an excess of a factor of 3 (instead
of 2.0\% for RGEs).

By using these results as reference we explored the properties of the $\sim$
300,000 galaxies in the SDSS/DR7 main sample, also dropping the constraint of
the stellar population age. Excluding the star-forming galaxies, the
distribution of EW turns out to be still strongly concentrated around
the values found for the old stellar population galaxies, but the fraction of
outliers is significantly higher (5.1 \% with residuals higher than a factor
of 5 and 6.6\% higher than 3). This is probably the result of the known
connection between star formation and nuclear activity
(e.g., \citealt{kauffmann03c}): the selection of regions dominated by an old
stellar population filters out a large part of AGN.

These results support the conclusion that stellar processes are generally
at the origin of the emission lines, while active galaxies are only a small
fraction of the overall galaxies population. Furthermore, the number of
galaxies with AGN-like emission line ratios is far higher (by a factor of
$\sim$ 5 - 10) than the fraction of outliers. The location of a galaxy in the
diagnostic diagrams is not a sufficient tool to establish the dominant source
of the ionizing photons, which is instead better predicted by the EW of the
emission lines, which in turn agrees with the conclusions of \citet{cid10b}.

Because the dominant contamination of the emission lines is apparently
associated with stars, a robust test for the reliability of the AGN census can
be performed considering spectra obtained on a smaller physical region. If the
emission lines are indeed mainly of nuclear origin, their contrast against the
continuum is expected to be enhanced, while for a dominant stellar origin no
significant changes are expected in terms of distribution of equivalent
width. And this latter is the result of our analysis, because the peak of
distribution of the ratio between line flux and continuum does not vary
significantly with redshift. A mild increase of the fraction of outliers is
instead observed at decreasing redshift. This suggests that in addition to the
dominant diffuse line emission, there is also a contribution from a nuclear
emission line region. In the more distant objects, this is generally swamped
by the stellar emission, while in nearer galaxies the presence of the AGN is
sufficient to increase the lines EW within the aperture significantly.

Even better data for this analysis come from the Palomar survey that covers
most bright nearby galaxies \citep{filippenko85}. The contribution of the
stellar emission in these spectra is reduced, with respect to the SDSS sample,
by a factor of 20-100. Nonetheless, the distribution of the ratio between line
and continuum is still strongly concentrated at low values. The Palomar survey
also confirms the trend of an AGN increase in nearer galaxies, because one
third of the objects in the AGN region also show a high EW, twice the fraction
of the SDSS galaxies with z$\sim0.01-0.02$.

We conclude that the current AGN census based on optical spectroscopy includes
a dominant fraction of objects where the emission lines are not powered by an
active nucleus, a problem particularly severe for the lowest luminosity
objects with a LINER spectrum. This has strong implications for our
understanding of AGN because a study based on a sample whose purity is only at
the $\sim$ 10\% level necessarily leads to spurious results. As a consequence
the properties of low-luminosity AGN should be fundamentally revised.

An improved census can be obtained from the spatially resolved optical
spectroscopy of the Palomar sample with a reduced aperture size, ideally with
integral field units of high spatial resolution. An improvement by a factor of
$\sim$10 from the 8 squared arcseconds of the original data is easily within
reach.

\acknowledgements

I thank Ranieri Baldi, David J. Axon, Roberto Rampazzo, and Francesca
Annibali for their comments and useful discussions.


\begin{thebibliography}{28}
\expandafter\ifx\csname natexlab\endcsname\relax\def\natexlab#1{#1}\fi

\bibitem[{{Annibali} {et~al.}(2010){Annibali}, {Bressan}, {Rampazzo},
  {Zeilinger}, {Vega}, \& {Panuzzo}}]{annibali10}
{Annibali}, F., {Bressan}, A., {Rampazzo}, R., {et~al.} 2010, A\&A, 519, A40+

\bibitem[{{Baldwin} {et~al.}(1981){Baldwin}, {Phillips}, \&
  {Terlevich}}]{baldwin81}
{Baldwin}, J.~A., {Phillips}, M.~M., \& {Terlevich}, R. 1981, \pasp, 93, 5

\bibitem[{{Bendo} {et~al.}(2010){Bendo}, {Wilson}, {Pohlen}, {Sauvage}, {Auld},
  {Baes}, {Barlow}, {Bock}, {Boselli}, {Bradford}, {Buat}, {Castro-Rodriguez},
  {Chanial}, {Charlot}, {Ciesla}, {Clements}, {Cooray}, {Cormier}, {Cortese},
  {Davies}, {Dwek}, {Eales}, {Elbaz}, {Galametz}, {Galliano}, {Gear}, {Glenn},
  {Gomez}, {Griffin}, {Hony}, {Isaak}, {Levenson}, {Lu}, {Madden},
  {O'Halloran}, {Okumura}, {Oliver}, {Page}, {Panuzzo}, {Papageorgiou},
  {Parkin}, {Perez-Fournon}, {Rangwala}, {Rigby}, {Roussel}, {Rykala},
  {Sacchi}, {Schulz}, {Schirm}, {Smith}, {Spinoglio}, {Stevens}, {Sundar},
  {Symeonidis}, {Trichas}, {Vaccari}, {Vigroux}, {Wozniak}, {Wright}, \&
  {Zeilinger}}]{bendo10}
{Bendo}, G.~J., {Wilson}, C.~D., {Pohlen}, M., {et~al.} 2010, \aap, 518, L65+

\bibitem[{{Binette} {et~al.}(1994){Binette}, {Magris}, {Stasi{\'n}ska}, \&
  {Bruzual}}]{binette94}
{Binette}, L., {Magris}, C.~G., {Stasi{\'n}ska}, G., \& {Bruzual}, A.~G. 1994,
  A\&A, 292, 13

\bibitem[{{Boquien} {et~al.}(2011){Boquien}, {Calzetti}, {Combes}, {Henkel},
  {Israel}, {Kramer}, {Rela{\~n}o}, {Verley}, {van der Werf}, \&
  {Xilouris}}]{boquien11}
{Boquien}, M., {Calzetti}, D., {Combes}, F., {et~al.} 2011, ArXiv e-prints

\bibitem[{{Buttiglione} {et~al.}(2010){Buttiglione}, {Capetti}, {Celotti},
  {Axon}, {Chiaberge}, {Macchetto}, \& {Sparks}}]{buttiglione10}
{Buttiglione}, S., {Capetti}, A., {Celotti}, A., {et~al.} 2010, \aap, 509, A6

\bibitem[{{Capetti} \& {Baldi}(2011)}]{capetti11alias}
{Capetti}, A. \& {Baldi}, R.~D. 2011, \aap, 529, A126+ (Paper I)

\bibitem[{{Cid Fernandes} {et~al.}(2011){Cid Fernandes}, {Stasi{\'n}ska},
  {Mateus}, \& {Vale Asari}}]{cid10b}
{Cid Fernandes}, R., {Stasi{\'n}ska}, G., {Mateus}, A., \& {Vale Asari}, N.
  2011, \mnras, 413, 1687

\bibitem[{{Filippenko} \& {Sargent}(1985)}]{filippenko85}
{Filippenko}, A.~V. \& {Sargent}, W.~L.~W. 1985, \apjs, 57, 503

\bibitem[{{Heckman}(1980)}]{heckman80}
{Heckman}, T.~M. 1980, A\&A, 87, 152

\bibitem[{{Ho} {et~al.}(1995){Ho}, {Filippenko}, \& {Sargent}}]{ho95}
{Ho}, L.~C., {Filippenko}, A.~V., \& {Sargent}, W.~L. 1995, ApJS, 98, 477

\bibitem[{{Ho} {et~al.}(1997){Ho}, {Filippenko}, \& {Sargent}}]{ho97}
{Ho}, L.~C., {Filippenko}, A.~V., \& {Sargent}, W.~L.~W. 1997, ApJS, 112, 315

\bibitem[{{Hyde} \& {Bernardi}(2009)}]{hyde09}
{Hyde}, J.~B. \& {Bernardi}, M. 2009, \mnras, 394, 1978

\bibitem[{{Kauffmann} {et~al.}(2003{\natexlab{a}}){Kauffmann}, {Heckman},
  {Tremonti}, {Brinchmann}, {Charlot}, {White}, {Ridgway}, {Brinkmann},
  {Fukugita}, {Hall}, {Ivezi{\'c}}, {Richards}, \& {Schneider}}]{kauffmann03c}
{Kauffmann}, G., {Heckman}, T.~M., {Tremonti}, C., {et~al.} 2003{\natexlab{a}},
  \mnras, 346, 1055

\bibitem[{{Kauffmann} {et~al.}(2003{\natexlab{b}}){Kauffmann}, {Heckman},
  {White}, {Charlot}, {Tremonti}, {Brinchmann}, {Bruzual}, {Peng}, {Seibert},
  {Bernardi}, {Blanton}, {Brinkmann}, {Castander}, {Cs{\'a}bai}, {Fukugita},
  {Ivezic}, {Munn}, {Nichol}, {Padmanabhan}, {Thakar}, {Weinberg}, \&
  {York}}]{kauffmann03b}
{Kauffmann}, G., {Heckman}, T.~M., {White}, S.~D.~M., {et~al.}
  2003{\natexlab{b}}, \mnras, 341, 33

\bibitem[{{Kewley} {et~al.}(2006){Kewley}, {Groves}, {Kauffmann}, \&
  {Heckman}}]{kewley06}
{Kewley}, L.~J., {Groves}, B., {Kauffmann}, G., \& {Heckman}, T. 2006, MNRAS,
  372, 961

\bibitem[{{Mulchaey} {et~al.}(1996){Mulchaey}, {Wilson}, \&
  {Tsvetanov}}]{mulchaey96}
{Mulchaey}, J.~S., {Wilson}, A.~S., \& {Tsvetanov}, Z. 1996, \apjs, 102, 309

\bibitem[{{Nakamura} {et~al.}(2003){Nakamura}, {Fukugita}, {Yasuda}, {Loveday},
  {Brinkmann}, {Schneider}, {Shimasaku}, \& {SubbaRao}}]{nakamura03}
{Nakamura}, O., {Fukugita}, M., {Yasuda}, N., {et~al.} 2003, \aj, 125, 1682

\bibitem[{{Oh} {et~al.}(2011){Oh}, {Sarzi}, {Schawinski}, \& {Yi}}]{oh11}
{Oh}, K., {Sarzi}, M., {Schawinski}, K., \& {Yi}, S.~K. 2011, ArXiv e-prints

\bibitem[{{Sandage} \& {Tammann}(1981)}]{sandage81}
{Sandage}, A. \& {Tammann}, G.~A. 1981, in Carnegie Inst. of Washington, Publ.
  635; Vol. 0; Page 0, 0--+

\bibitem[{{Sarzi} {et~al.}(2010){Sarzi}, {Shields}, {Schawinski}, {Jeong},
  {Shapiro}, {Bacon}, {Bureau}, {Cappellari}, {Davies}, {de Zeeuw}, {Emsellem},
  {Falc{\'o}n-Barroso}, {Krajnovi{\'c}}, {Kuntschner}, {McDermid}, {Peletier},
  {van den Bosch}, {van de Ven}, \& {Yi}}]{sarzi10}
{Sarzi}, M., {Shields}, J.~C., {Schawinski}, K., {et~al.} 2010, \mnras, 402,
  2187

\bibitem[{{Shen} {et~al.}(2003){Shen}, {Mo}, {White}, {Blanton}, {Kauffmann},
  {Voges}, {Brinkmann}, \& {Csabai}}]{shen03}
{Shen}, S., {Mo}, H.~J., {White}, S.~D.~M., {et~al.} 2003, \mnras, 343, 978

\bibitem[{{Stasi{\'n}ska} {et~al.}(2008){Stasi{\'n}ska}, {Vale Asari}, {Cid
  Fernandes}, {Gomes}, {Schlickmann}, {Mateus}, {Schoenell}, \&
  {Sodr{\'e}}}]{stasinska08}
{Stasi{\'n}ska}, G., {Vale Asari}, N., {Cid Fernandes}, R., {et~al.} 2008,
  MNRAS, 391, L29

\bibitem[{{Strauss} {et~al.}(2002){Strauss}, {Weinberg}, {Lupton}, {Narayanan},
  {Annis}, {Bernardi}, {Blanton}, {Burles}, {Connolly}, {Dalcanton}, {Doi},
  {Eisenstein}, {Frieman}, {Fukugita}, {Gunn}, {Ivezi{\'c}}, {Kent}, {Kim},
  {Knapp}, {Kron}, {Munn}, {Newberg}, {Nichol}, {Okamura}, {Quinn}, {Richmond},
  {Schlegel}, {Shimasaku}, {SubbaRao}, {Szalay}, {Vanden Berk}, {Vogeley},
  {Yanny}, {Yasuda}, {York}, \& {Zehavi}}]{strauss02}
{Strauss}, M.~A., {Weinberg}, D.~H., {Lupton}, R.~H., {et~al.} 2002, \aj, 124,
  1810

\bibitem[{{Tremaine} {et~al.}(2002){Tremaine}, {Gebhardt}, {Bender}, {Bower},
  {Dressler}, {Faber}, {Filippenko}, {Green}, {Grillmair}, {Ho}, {Kormendy},
  {Lauer}, {Magorrian}, {Pinkney}, \& {Richstone}}]{tremaine02}
{Tremaine}, S., {Gebhardt}, K., {Bender}, R., {et~al.} 2002, \apj, 574, 740

\bibitem[{{Trinchieri} \& {di Serego Alighieri}(1991)}]{trinchieri91}
{Trinchieri}, G. \& {di Serego Alighieri}, S. 1991, \aj, 101, 1647

\bibitem[{{Veilleux} \& {Osterbrock}(1987)}]{veilleux87}
{Veilleux}, S. \& {Osterbrock}, D.~E. 1987, \apjs, 63, 295

\bibitem[{{York} {et~al.}(2000){York}, {Adelman}, {Anderson}, {Anderson},
  {Annis}, {Bahcall}, {Bakken}, {Barkhouser}, {Bastian}, {Berman}, {Boroski},
  {Bracker}, {Briegel}, {Briggs}, {Brinkmann}, {Brunner}, {Burles}, {Carey},
  {Carr}, {Castander}, {Chen}, {Colestock}, {Connolly}, {Crocker}, {Csabai},
  {Czarapata}, {Davis}, {Doi}, {Dombeck}, {Eisenstein}, {Ellman}, {Elms},
  {Evans}, {Fan}, {Federwitz}, {Fiscelli}, {Friedman}, {Frieman}, {Fukugita},
  {Gillespie}, {Gunn}, {Gurbani}, {de Haas}, {Haldeman}, {Harris}, {Hayes},
  {Heckman}, {Hennessy}, {Hindsley}, {Holm}, {Holmgren}, {Huang}, {Hull},
  {Husby}, {Ichikawa}, {Ichikawa}, {Ivezi{\'c}}, {Kent}, {Kim}, {Kinney},
  {Klaene}, {Kleinman}, {Kleinman}, {Knapp}, {Korienek}, {Kron}, {Kunszt},
  {Lamb}, {Lee}, {Leger}, {Limmongkol}, {Lindenmeyer}, {Long}, {Loomis},
  {Loveday}, {Lucinio}, {Lupton}, {MacKinnon}, {Mannery}, {Mantsch}, {Margon},
  {McGehee}, {McKay}, {Meiksin}, {Merelli}, {Monet}, {Munn}, {Narayanan},
  {Nash}, {Neilsen}, {Neswold}, {Newberg}, {Nichol}, {Nicinski}, {Nonino},
  {Okada}, {Okamura}, {Ostriker}, {Owen}, {Pauls}, {Peoples}, {Peterson},
  {Petravick}, {Pier}, {Pope}, {Pordes}, {Prosapio}, {Rechenmacher}, {Quinn},
  {Richards}, {Richmond}, {Rivetta}, {Rockosi}, {Ruthmansdorfer}, {Sandford},
  {Schlegel}, {Schneider}, {Sekiguchi}, {Sergey}, {Shimasaku}, {Siegmund},
  {Smee}, {Smith}, {Snedden}, {Stone}, {Stoughton}, {Strauss}, {Stubbs},
  {SubbaRao}, {Szalay}, {Szapudi}, {Szokoly}, {Thakar}, {Tremonti}, {Tucker},
  {Uomoto}, {Vanden Berk}, {Vogeley}, {Waddell}, {Wang}, {Watanabe},
  {Weinberg}, {Yanny}, \& {Yasuda}}]{york00}
{York}, D.~G., {Adelman}, J., {Anderson}, Jr., J.~E., {et~al.} 2000, \aj, 120,
  1579

\end{thebibliography}
\end{document}